\documentstyle[12pt,epsfig,rotating]{article}
\def\pge{\pagestyle{empty}} \def\pgn{\pagestyle{plain}}
\def\bsls35{\baselineskip 0.35in}
     
\def\spg{\setcounter{page}} 
\def\bd{
\begin{document}} \def\ed{\end{document}}
\def\bmp{\begin{minipage}} \def\emp{\end{minipage}}
\def\bcc{\begin{center}} \def\ecc{\end{center}}     \def\npg{\newpage}
\def\beq{\begin{equation}} \def\eeq{\end{equation}} \def\hph{\hphantom}
\def\be{\begin{equation}} \def\ee{\end{equation}} \def\r#1{$^{[#1]}$}
\def\n{\noindent} \def\ni{\noindent} \def\pa{\parindent} 
\def\hs{\hskip} \def\vs{\vskip} \def\hf{\hfill} \def\ej{\vfill\eject} 
\def\cl{\centerline} \def\ob{\obeylines}  \def\ls{\leftskip}
\def\underbar#1{$\setbox0=\hbox{#1} \dp0=1.5pt \mathsurround=0pt
   \underline{\box0}$}   \def\ub{\underbar}    \def\ul{\underline} 
\def\f{\left} \def\g{\right} \def\e{{\rm e}} \def\o{\over} 
\def\vf{\varphi} \def\pl{\partial} \def\cov{{\rm cov}} \def\ch{{\rm ch}}
\def\la{\langle} \def\ra{\rangle} \def\EE{e$^+$e$^-$}
\def\bitz{\begin{itemize}} \def\eitz{\end{itemize}}
\def\btbl{\begin{tabular}} \def\etbl{\end{tabular}}
\def\btbb{\begin{tabbing}} \def\etbb{\end{tabbing}}
\def\beqar{\begin{eqnarray}} \def\eeqar{\end{eqnarray}}
\def\\{\hfill\break} \def\dit{\item{-}} \def\i{\item} 
\def\bbb{\begin{thebibliography}{9} \itemsep=-1mm}
\def\ebb{\end{thebibliography}} \def\bb{\bibitem}
\def\bpic{\begin{picture}(260,240)} \def\epic{\end{picture}}
\def\akgt{\noindent{\bf Acknowledgements}}
\def\fgn{\noindent{\bf\large\bf Figure captions}}
\def\pt{{p_{\rm t}}} \def\vf{\varphi} \def\yct{y_{\rm cut}} 
%%%%%%%%%%%%%%%%%%%%%%%%%%%%%%%%%%%%%%%%%%%%%%%%%%%%%%%%%%%%%%%%%%%%
\bd
\pge

\null{}\vskip -1.2cm
\hskip12cm{\bf HZPP-0003}
%%%%%%%%%%%%%%%%%%%%%%%%%%%  hep-ph/0002244 (855ay)
%%%%%%%%%%%%%%%%%%%%%%%%%%%  LP7638
\vskip-0.2cm

\hskip12cm Feb. 25, 2000

\vskip1cm

\begin{center}
{\Large Anisotropy of Dynamical Fluctuations as a Probe
\vskip0.4cm

for Soft and Hard Processes in High Energy 
Collisions\footnote{This work is supported in part by the NSFC 
under project 19975021.}}
\vskip0.5cm

\vskip0.2cm

{\large Liu Lianshou,  \ \  Chen Gang  \ \ and \ \ Fu Jinghua}
%%%%{\large    Liu Lianshou \ \ and \ \ Chen Gang}

{\small Institute of Particle Physics, Huazhong Normal University,
Wuhan 430079 China}
\date{ }
\end{center}

\begin{center}
\begin{minipage}{125mm}
\vskip 0.5in
\begin{center}{\Large ABSTRACT}\end{center}
{\hskip0.6cm
It is shown using Lund Monte Carlo that, unlike the
average properties of the hadronic system inside jets,
the anisotropy of dynamical fluctuations in these systems
changes abruptly with the variation of the cut parameter $\yct$.
A transition point exists, where the dynamical fluctuations in the
hadronic system inside jet behave like those in soft hadronic collisions.
Thus the anisotropy property of the dynamical fluctuations can serve as
a probe for the soft and hard processes in high energy collisions.}
\end{minipage}
\end{center}
\vs0.8cm

{\large PACS number: 13.85 Hd

\vs0.5cm
\ni
\null{}\hskip2.6cm processes}

%%% \hskip1.8cm
%%% \ hard and soft processes}

\npg \pgn \spg{2}
\baselineskip 0.24in

As is well known, the presently most promissing theory of strong interaction
--- Quantum Chromo-Dynamics (QCD) has the special property of both asymptotic
freedom and colour confinement. For this reason, in any process, even though 
the energy scale, $Q^2$, is large enough to be able to do perturbative QCD 
(pQCD) calculation, there must be a non-perturbative hadronization phase 
before the final state particles can be observed. Therefore, the transition 
or interplay between hard and soft processes is a very important problem.

In current literature, this transition is determined by some cut-parameter.
For example, in doing theoretical calculation a parameter $Q_0^2$ is 
introduced. 
When $Q^2>Q_0^2$ the  perturbative QCD is assumed to be applicable and the 
process is hard. While when $Q^2<Q_0^2$ the perturbative calculation is 
unallowed and the process becomes soft (nonperturbative). However, the 
value of $Q_0^2$ is not determined exactly. It decreases steadily as
the developement of perturbative technique.
%%% can be chosen within a certain range.

In experimental data analysis people use some ``jet-algorithm'' (e.g.
Jade~\cite{Jade} or Durham~\cite{Durham} ones) to combine 
the final-state particles into ``jets''.
Each jet is assumed to be originated from a hard parton, and the hadrons in
the jet is produced softly from this hard parton. Thus the 
transition between hard and soft processes is described as the production
of hard partons and the subsequent hadronization of these partons. In this
formalism there is also a parameter --- $\yct$. The value of this parameter
determines how the hadrons are grouped into jets, and whether an event is 
a ``2-jet event'' or a ``3-jet'', ``4-jet'' ones,
%%% Through changing the value 
%%% of $\yct$, a ``2-jet event'' may become a ``3-jet'' or even ``4-jet'' ones,
%%% and vice versa. 
%%% There is no unique way to determine which value of $\yct$
%%% is more reasonable.

Let us concentrate on the 2-jet events. By definition, these two jets
should be developed softly from two hard partons and no hard process is
involved in the evolution. If there is any hard process in the 
developement then we say that a third jet appears.  Historically, 
it was the observation of the third jet in e$^+$e$^-$ collisions that 
confirmed the existence of gluon~\cite{Brandelik}--\cite{Bartel}. 
In this sense, there should be a definite value of $\yct$, which is 
consistent with the physical meaning of ``jet''. 

On the other hand, due to the success in pQCD calculation of jet, people
sometimes take the number of jets in an event as indefinite, depending on
the value of $\yct$, which can be chosen arbitrarily. Their stress
is in ultilizing this dependence to confront the pQCD calculation
with experiments.  From this point of view, the physical meaning of jet 
and the associated concepts ------ ``soft'' and ``hard'' are neglected. 
A process is hard or soft is not judged physically, 
but is determined through the technical problem
of whether the process can be calculated by perturbative QCD.

Let us remind that
physically, soft and hard are distinguished through the magnitude of
transverse momentum. In hadron-hadron collisions at energies below
top-ISR most of the final-state hadrons have low transverse momenta
and the process is soft. At collider energies high-transverse-momentum
jets, coming from hard parton collisions, start to appear~\cite{UA1jet}.
The transverse momenta of this jets are higher than 10 -- 20 GeV/$c$.
Besides, there are also mini-jets with transverse momenta higher than
about 4 -- 5 GeV/$c$~\cite{Ciapetti}, which are generally refered to as
semi-hard.  The critical value of transverse momentum for the transition
between soft and hard (semi-hard) is about 4 -- 5 GeV/$c$.

The following important questions arise: 
1) Does the number of jets in an event possess any definite meaning?
If yes, how to determine this number, i.e. how to decide the correct 
value of $\yct$ for the determination of this number.
2) Is it in principle possible to locate
the transition between soft and hard processes in the hadronic final
states of high energy e$^+$e$^-$ collisions? If yes, how to do that?

In order to answer these questions, let us remind that
the qualitative difference between the typical soft process ---  moderate 
energy hadron-hadron collisions and the typical hard process ---
high energy e$^+$e$^-$ collisions can be observed most clearly 
in the property of dynamical fluctuations therein.
It is found recently~\cite{FFLPRD} 
that inspite of the similarities in the average
properties, the dynamical fluctuations in the hadronic systems from
these two processes are qualitatively different ------ the former is
anisotropic in the longitudinal-transvere plane and isotropic in the
transverse planes while the latter is isotropic in three dimensional phase 
space.

This observation inspired us to think that the dynamical-fluctuation property
may provide a probe for the transition betweem soft and hard processes
inside the hadronic final state of high energy e$^+$e$^-$ collisions. In
the present letter we show, using Lund Monte Carlo simulation, that this is 
indeed the case.  

In total 500 000 events are generated for 91.2 GeV e$^+$e$^-$ collisions using 
JETSET7.4. The resulting hadronic systems are analysed using Durham 
and/or Jade jet-algorithms. The fractions $R_2$, $R_3$, $R_4$ of the 
2-, 3-, 4-jet events in the whole sample are plotted versus the value
of $\yct$ for both Durham and Jade algorithms in Fig.1. It can be seen 
clearly form the figures that  the definition of ``jet'' depends strongly on 
the value of $\yct$. When $\yct$ is big, most of the events are taken to be 
``2-jet'' events. In the limit of very large $\yct$, the whole sample consists 
of only ``2-jet'' events. On the contrary, when the value of $\yct$ decreases 
continuously, the jets are divided further and further, and gradually most of
the events become ``multi-jet'' (more than two jets) ones.   

At the energy in consideration, it is certainly impossible that all the 
events are 2-jet ones.
Neither is it possible that most of the events are multi-jet ones. In order
to determine a reasonable value of $\yct$, we have to use the dependence of
some physical property of the system on $\yct$. As example, we show in Fig.2 
the dependence of average charged
multiplicity $\la N_{\rm ch}\ra$ and average ellipticity $\la e\ra$ on $\yct$
for the ``2-jet'' sample determined by Durham algorithm. The ellipticity $e$
is an event-shape parameter defined as the ratio of minor $T_3$ to major
$T_2$ in thrust analysis~\cite{Brandt}\cite{Barber}
\beq  %% 1
 e = T_3 / T_2.
\eeq
By definition $e\leq 1$.  When $e=1$ the jet cone is circular in the momentum 
space. It is expected that, when $\yct$ increases, more and more ``inpurities''
(multi-jet events) are mixed into the ``2-jet'' event sample, and the 
jet cone will diviate more and more from being circular. So the average
ellipticity $\la e\ra$ will drcrease with the increasing of $\yct$.
It can be seen from the figure that this is indeed the case. However, the 
value of $\la e\ra$ changes smoothly with $\yct$ and it is hard to get a 
probe for a reasonable value of $\yct$ by using $\la e\ra$. The same holds 
also for $\la N_{\rm ch}\ra$ and other average quantities.

\begin{center}
\begin{picture}(250,450)
\put(-120,290)
{
{\epsfig{file=fig1.epsi,width=225pt,height=120pt}}
}
\put(170,290)
{
{\epsfig{file=fig2.epsi,width=170pt,height=175pt}}
}
\end{picture}
\end{center}

\vs-10.5cm
\n{\small Fig.1 \ \ The ratio of 2-, 3-, 4-jet events \hskip2cm Fig.2 \ \
Average charged multiplicity and ellipticity}

{\small as function of $\yct$ \hskip5cm as function of $\yct$}
\vs0.5cm

Let us turn now to consider the dynamical fluctuations. These fluctuations
can be characterized by the anomalous scaling of 
factorial moments (FM)~\cite{BP}:
\beqar   %%% (2)
  F_q(M)&=&{\frac {1}{M}}\sum\limits_{m=1}^{M}{{\langle n_m(n_m-1)
     \cdots (n_m-q+1)\rangle }\over {{\langle n_m \rangle}^q}}\\ \nonumber
   &\propto& (M)^{\phi_q}\ \  \quad \quad (M\to \infty) \ \ ,
\eeqar
where a region $\Delta$ in 1-, 2- or 3-dimensional phase space is
divided into $M$ cells, $n_m$  is the multiplicity in the $m$th cell,
and $\langle\cdots\rangle$ denotes vertically averaging over the event
sample. Note that when the fluctuations exist in higher-dimensional (2-D
or 3-D) space the projection effect~\cite{Ochs} will cause the 
second-order 1-D FM goes
to saturation according to the rule\footnote{In order to elliminate the 
influence of momentum conservation~\cite{MMTN}, the first few points 
($M=1,2$ or 3) should be omitted when fitting the data to Eq.(3).}:
\be   %%% (3)
 F_2^{(a)}(M_a) = A_a-B_a M_a^{-\gamma_a}, \ \ 
\ee
where $a=1,2,3$ denotes the different 1-D variables. The parameter $\gamma_a$
describes the rate of going to saturation of the FM in direction $a$ and is 
the most
important characteristic for the higher-dimensional dynamical fluctuations.
If $\gamma_a = \gamma_b$ the fluctuations are isotropic in the $a,b$
plane; while when $\gamma_a \neq \gamma_b$ the fluctuations are anisotropic 
in this plane. The degree of anisotropy is characterized by the Hurst
exponent $H_{ab}$, which can be obtained from the values of $\gamma_a$
and $\gamma_b$ as~\cite{ZGKX}
\be   %%% (4)
 H_{ab} = {1+\gamma_b\over 1+\gamma_a}.
\ee
The dynamical fluctuations are isotropic when  $H_{ab} = 1$, and anisotropic 
when $H_{ab} \neq 1$.

For the 250 GeV/$c$ $\pi$(K)-p collisions from NA22 the Hurst 
exponents are found to be~\cite{NA22}:
\be %%% (5)
H_{\pt\vf}=0.99 \pm 0.01, \ \ 
 H_{y\pt}=0.48 \pm 0.06, \ \ H_{y\vf}=0.47 \pm 0.06,  
\ee
which means that the dynamical fluctuations in this moderate energy
hadron-hadron collisions are isotropic in the
transverse plane and anisotropic in the longitudinal-transvere planes.
This is what should be~\cite{WLprl}, because there is almost no hard 
collisions at
this energy and the direction of motion of the incident hadrons 
(longitudinal direction) should be previleged. Note that the special
role of longitudinal direction in these soft processes is present
both in the magnitude of average momentum and in the dynamical fluctuations 
in phase space.

In high energy e$^+$e$^-$ collisions, the longitudinal direction is chosen 
along the thrust axis, which is the direction of motion of the primary
quark-antiquark pair. Since this pair of quark and antiquark move back to back
with very high momenta, the magnitude of average momentum of final state 
hadrons is also anisotropic due to momentum conservation.  
However, the dynamical fluctuations in this case come from 
the QCD branching of partons~\cite{Vineziano}, which is isotropic
in nature. Therfore, although the momentum distribution still has 
an elongated shape, the dynamical fluctuations in this case should be 
isotropic in 3-D phase space.

A Monte Carlo study for e$^+$e$^-$ collisions at 91.2 GeV confirms this
assertion~\cite{FFLPRD}. The dynamical fluctuations are 
approximately isotropic in the 
3-D phase space, the corresponding Hurst exponents being 
\be %%% (6)
H_{\pt\vf}=1.18 \pm 0.03, \ \ 
 H_{y\pt}=0.95 \pm 0.02, \ \  H_{y\vf}=1.11 \pm 0.02. 
\ee
The present available experimental data for e$^+$e$^-$ collisions at 91.2 GeV 
also show isotropic dynamical fluctuations in 3-D~\cite{DELPHI}.

Now we apply this technique to the ``2-jet'' sample obtained from a 
certain, e.g. Durham, jet-algorithm with some definite value of $\yct$. 
Doing the analysis for different values of $\yct$,
the dependence of dynamical-fluctuation property of the ``2-jet'' 
sample on the value of $\yct$ can be investigated.

Let us try to discuss what results can be expected?

As we have shown in Fig.1, when $\yct$ is very big the ``2-jet'' sample 
coincides with the whole event sample, $R_2 = 1$. In this case, the 
fluctuations are 
known to be isotropic in the 3-D phase space, cf. Eq.(6), i.e. the parameter 
$\gamma_a$ for the three 1-D variables ($y,\pt,\vf$) equal to each other
($\gamma_\pt = \gamma_\vf = \gamma_y$). 

\begin{center}
\begin{picture}(250,450)
\put(-10,250)
{
{\epsfig{file=fig3.epsi,width=240pt,height=200pt}}
}
\end{picture}
\end{center}

\vs-9cm
\cl{($a$)\hskip4cm ($b$)}

\cl{\small Fig.3 \ \ The variation of $\gamma$ with $R_2$ and $\yct$}

As the decreasing of $\yct$ the multi-jet events, which contaminate the 
``2-jet'' sample, will be cleared away gradually, and at a certain value 
of $\yct$, a ``pure'' 2-jet sample will be formed.  The word ``pure'' is 
used here to indicate that these two jets are developed softly from initial 
partons and no other jet(s) has been mixed in.

It can be expected that the dynamical fluctuations in the ``pure'' 2-jet
sample will mimic those in the soft hadronic collisions, i.e.
isotropic in the transverse plane and anisotropic in the 
longitudinal-transverse planes ($\gamma_\pt = \gamma_\vf \neq \gamma_y$).

Thus the variation of $\gamma$'s with the decreasing of $\yct$ (or
decreasing of $R_2$) is expected to be: 
At first, when $\yct$ is very big, the ``2-jet'' sample is identical to
the whole event sample ($R_2=1$), and the three
$\gamma$'s equal to each other; As the decreasing of $\yct$ (the
deceasing of $R_2$)
the three $\gamma$'s depart, and becomes, at a certain value of $\yct$,
isotropic in ($\pt,\vf$) and anisotropic in ($y,\pt$) and ($y,\vf$),
$\gamma_\pt = \gamma_\vf \neq \gamma_y$. 

The results of simulation are presented in Fig.3($a$). It can be seen from
the figure that the above expectation comes true. The characteristic
behaviour $\gamma_\pt = \gamma_\vf \neq \gamma_y$ arrives at
$\yct\approx 0.0048$ ($R_2\approx 0.48$).  The values of $\gamma$'s 
and the corresponding Hurst exponents at this point are listed in Table I.
For convenience we will call this point, where $\gamma_\pt = \gamma_\vf 
\neq \gamma_y$, as transition point.

\vs0.5cm
\cl{Table I Parameter $\gamma$ and Hurst exponents at the transition point}

\def\bcc{\begin{center}} \def\ecc{\end{center}}
\def\btbl{\begin{tabular}} \def\etbl{\end{tabular}}
\bcc\btbl{|c|c|c|c|c|c|}\hline
\multicolumn{6}{|c|}{$\yct=0.0048$ (Durham) \qquad \qquad
$R_{\rm 2jet}=0.48$} \\ \hline
$\gamma_y$ & $\gamma_\pt$ & $\gamma_\vf$ & 
$H_{y\pt}$ & $H_{y\vf}$ & $H_{\pt\vf}$ \\ \hline 
1.074$\pm$0.037 & 0.514$\pm$0.080 & 0.461$\pm$0.021&
0.73$\pm$0.06 & 0.70$\pm$0.06& 
 0.96$\pm$0.10 \\ \hline
\etbl\ecc

\begin{center}
\begin{picture}(250,450)
\put(30,280)
{
{\epsfig{file=fig4.epsi,width=180pt,height=170pt}}
}
\end{picture}
\end{center}

\vs-10.5cm
\cl{\small Fig.4 \ \ Comparison of the speed of going to saturation of}

\cl{\small
$F_2$ for different 1-D variables at different $R_2$}
\vs0.5cm

Note that in the Durham algorithm that we are using the test variable $y$ 
is essentially the relative transverse 
momentum $k_\perp$ squared~\cite{Dokshitzer}. 
The transition point $\yct\approx 0.0048$ corresponds to 
$k_\perp \approx 4$ GeV/c, which 
is consistent with the critical value of transverse momentum between
soft and hard (semi-hard) components in hadron-hadron collisions.

It is instructive also to follow the evolusion of $\gamma$'s with the 
increasing of $\yct$ (incresing of $R_2$).

It can be seen from Fig.3($a$) that, when $\yct$ ($R_2$) is very small, 
where the two ``jets'' are highly undeveloped and each consists mainly 
of one hard parton,
$\gamma_\vf$ is consistent to zero, i.e. there is no dynamical fluctuation
in $\vf$ at all. On the other hand, at this point $\gamma_\pt$ is almost as 
large as $\gamma_y$, showing that
the dynamical fluctuations in this undeveloped ``2-jet'' system behaves
as an isotropic 2-D fractal in the ($y,\pt$) plane. 

When $\yct$ ($R_2$) increases, $\gamma_\pt$ departs with $\gamma_y$ and 
approaches to $\gamma_\vf$. What is important is that  $\gamma_\pt$ and
$\gamma_\vf$, instead of going up parallelly, cross over each other, turns 
from $\gamma_\vf < \gamma_\pt$ to  $\gamma_\vf > \gamma_\pt$, 
resulting in a sharp transition point. 
After that, the three $\gamma$'s approach eventually to a common value,
and the ``2-jet'' sample approachs to the whole event sample.

In order to show the evolusion of the
anisotropy property of dynamical fluctuations
with the variation of $\yct$ ($R_2$) more clearly, we take three typical 
points: $(A)$ \ $R_2=0.18$, $(B)$ \  $R_2=0.48$, $(C)$  \ $R_2=1$, indicated by 
arrows in Fig.3($a$). Point $A$ corresponds to the case of undeveloped jets, 
$B$ is the transition point and $C$ is the whole sample. Since the anisotropy
property of dynamical fluctuations determines solely by the rate of 
approaching to saturation of FM, which is characterized by the parameter 
$\gamma$, we rescale the $F_2(\pt)$ and $F_2(\vf)$ appropriately, letting them
coincide with $F_2(y)$ at $M=3$ and arrive at a common saturation height with
$F_2(y)$.  The results are shown in Fig.4. It can be seen from the figure that 
when $R_2=0.18$, $F_2(\vf)$ does not increase with $M$, i.e. no dynamical 
fluctuation at all in $\vf$, while at this point $F_2(\pt)$ and $F_2(y)$ 
go to saturation almost
with the same speed. When $R_2=0.48$ (transition point), $F_2(\pt)$ and
$F_2(\vf)$ go to saturation almost with the same speed, much slower than
$F_2(y)$ do. When $R_2=1$ (whole sample) all three $F_2$ coincide and go
to saturation with an identical speed.

For comparison, we have also done the same analysis using Jade algorithm.
The results, shown in Fig.3($b$), are qualitatively the same:
At small $\yct$ ($R_2$), $\gamma_\vf$ vanishes and $\gamma_\pt \approx
\gamma_y$; As $\yct$ ($R_2$) increases $\gamma_\pt$ and $\gamma_\vf$ 
approaches each other and cross over at $\yct \approx 0.158$ 
($R_2 \approx 0.39$). This is the transition point for Jade algorithm.
The parameter $\gamma$'s at this point are
$\gamma_y=1.22\pm 0.04$, $\gamma_\pt=0.51\pm 0.09$, $\gamma_\vf=0.59\pm 0.08$. 

In this letter we have shown using Lund Monte Carlo that, unlike the 
smooth change of average properties of the hadronic system inside jets, 
the anisotropy of dynamical fluctuations in these systems
changes abruptly with the variation of the cut parameter $\yct$. At
$\sqrt s = 91.2$ GeV, the dynamical fluctuations in the whole e$^+$e$^-$
collision sample (large $\yct$ limit) are fully isotropic in the 3-D phase 
space, and become highly anisotropic (almost no fluctuation at all in 
$\vf$) for small $\yct$ where the ``jet'' is highly undeveloped. A 
transition point exists, where the hadronic system inside jet
behaves like that of the soft hadronic collisions, i.e. the dynamical
fluctuations are isotropic in the transverse plane and anisotropic in the
longitudinal-transverse planes. The corresponding relative transverse momentum 
at the transition point is about $k_\perp \approx 4$ GeV/c, which
is consistent with the critical value of transverse momentum between
soft and hard (semi-hard) components in hadron-hadron collisions.  
Thus the the transition point determines the physically meaningful value 
of $\yct$, and thereby gives the number of jets in an events. The 
anisotropy property of the dynamical fluctuations can serve as 
a sensible probe for hard and soft processes.

This observation is not only meaningful in the study of jets in e$^+$e$^-$
collisions but also enlightening in the jet-physics in relativistic
heavy ion collisions, which will become important~\cite{WXN} after the
operation of the new generation of colliders at BNL (RHIC) and CERN (LHC).
 
\vskip0.5cm

\n{\bf Acknoledgement} 

%%% \n The authors are grateful to Fu Jinghua, Wu Yuanfang
\n The authors are grateful to Wu Yuanfang
and Xie Qubin for valuable discussions.

\vskip1.5cm
%%%%%%%%%%%%%%%%%%%%%%%%%%%%%%%%%%%%%%%%%%%%%%%%%%%%%%%%%%%%%%%%%%%%%%
%%% \def\Journal#1#2#3#4{{#1} {\bf #2}, #3 (#4)}
\def\Journal#1#2#3#4{{#1} {\bf #2} (#4) #3}
\def\NCA{\em Nuovo Cimento} \def\NIM{\em Nucl. Instrum. Methods}
\def\NIMA{{\em Nucl. Instrum. Methods}| {\bf A}}
\def\NPB{{\em Nucl. Phys.} {\bf B}}
\def\PLB{{\em Phys. Lett.} {\bf B}} \def\PRL{\em Phys. Rev. Lett.}
\def\PRD{{\em Phys. Rev.} {\bf D}} \def\ZPC{{\em Z. Phys.} {\bf C}}
%%%%%%%%%%%%%%%%%%%%%%%%%%%%%%%%%%%%%%%%%%%%%%%%%%%%%%%%%%%%%%%%%%%%%%

\ed
\newpage

\ni{\Large\bf Figure Captions}
\vskip0.8cm

\n{\bf Fig.1} \ \ The ratio of 2-, 3-, 4-jet events 
as function of $\yct$
\vs0.5cm

\n{\bf Fig.2} \ \ Average charged multiplicity and ellipticity
as function of $\yct$
\vs0.5cm

\n{\bf Fig.3} \ \ The variation of $\gamma$ with $R_2$ and $\yct$
\vs0.5cm

\n{\bf Fig.4} \ Comparison of the speed of going to saturation of
$F_2$ for different 1-D variables at different $R_2$

\newpage
\begin{center}
\begin{picture}(250,450)
\put(5,290)
{
{\epsfig{file=figure/ps/fig1.epsi,width=225pt,height=120pt}}
}
\put(35,5)
{
{\epsfig{file=figure/ps/fig2.epsi,width=170pt,height=175pt}}
}
\end{picture}
\end{center}

\vs-9.5cm
\cl{Fig.1}

\vs9.5cm
\cl{Fig.2}

\newpage

\begin{center}
\begin{picture}(250,450)
\put(-70,60)
{
{\epsfig{file=figure/ps/fig3.epsi,width=360pt,height=300pt}}
%%% {\epsfig{file=figure/ps/fig3.epsi,width=240pt,height=200pt}}
}
\end{picture}
\end{center}

\vs-2cm
\cl{($a$)\hskip4cm ($b$)}

\cl{Fig.3}
\newpage

\begin{center}
\begin{picture}(250,450)
\put(-30,130)
{
{\epsfig{file=figure/ps/fig4.epsi,width=270pt,height=255pt}}
%%% {\epsfig{file=figure/ps/fig4.epsi,width=180pt,height=170pt}}
}
\end{picture}
\end{center}

\vs-2.5cm
\cl{Fig.4}